\documentclass[conference, letterpaper]{IEEEtran}
\IEEEoverridecommandlockouts
\usepackage{cite}
\usepackage{amsmath,amssymb,amsfonts}
\usepackage{algorithm,algorithmic}
\usepackage{graphicx}
\usepackage{mathtools}
\usepackage{textcomp}
\usepackage{xcolor}
\usepackage{subcaption}
\usepackage{tabularx,booktabs}
\usepackage{multirow}
\usepackage{hyperref}

\usepackage{caption}
\usepackage{setspace}
\def\BibTeX{{\rm B\kern-.05em{\sc i\kern-.025em b}\kern-.08em
    T\kern-.1667em\lower.7ex\hbox{E}\kern-.125emX}}
\begin{document}

\title{Discrete-Time State-Feedback Controller with Canonical Form on Inverted Pendulum (on a cart)\\
}

\author{\IEEEauthorblockN{Bambang L Widjiantoro\IEEEauthorrefmark{1}, and
Moh Kamalul Wafi\IEEEauthorrefmark{2}}
\IEEEauthorblockA{\IEEEauthorrefmark{1,2}Engineering Physics Department, Institut Teknologi Sepuluh Nopember, Indonesia, 60111}
(\IEEEauthorrefmark{1}blelono@its.ac.id, \IEEEauthorrefmark{2}kamalul.wafi@its.ac.id) 
}


\maketitle

{\small \noindent \textit{\textbf{Abstract-}} The scope of inverted pendulum has been widely studied as one of the notable research with respect to standing in balance. The concept of this pendulum is similar to missile guidance, meaning that the center of drag is ahead that of gravity. Mathematical model of inverted pendulum on a cart is moreover presented in this paper. Various rewarding parameters are proposed from the displacement of the pivot, angular rotation, to external force exerted on the carriage so as to gain its equilibrium points and the linearized systems. Due to the severe risk of instability, a reliable closed-loop state feedback controller is designed to stabilize in upright position, even with large deviations. The specific concept proposed is to apply the canonical form of computing the determinant of gain $K$ leading to $K_d$. The results show that the constructed design can maintain the stability of the system by applying three sorts of initial condition and choosing sampling time $T$ under $0.2$ with small possible degrading performance.\\
\noindent \textit{\textbf{Keywords-}} \textit{State-Feedback}, \textit{Discrete-Time Control}, \textit{Controllable Canonical Form}, \textit{Inverted Pendulum}}

\section{Introduction}
The interest of inverted pendulum has increasingly been popular from the various control problem and its historical scope has been widely studied by \cite{Ref1}. The further concept of the pendulum is the wheeled inverted pendulum leading to divergent applications, such as in robotic and vehicle with adaptive self-dynamical balancing \cite{Ref2}, anti-slip balancing control \cite{Ref3} and various optimal controls \cite{Ref4}. Indeed, the stability of the pendulum lies in the mathematical theory as presented in \cite{Ref5} using partial differential equation (PDE) with parametrization in a larger class of port-controlled Hamiltonian. Furthermore, the appropriate control \cite{Ref6}, the design of the trajectory tracking \cite{Ref7} along with some optimizations \cite{Ref8} affect the equilibrium of the pendulum. However, those ideas always coincide with the complexity of the non-linear system so that the swing-up control applying cascade scheme is required \cite{Ref9}, \cite{Ref10} and it needs to be implemented in the real environment as in \cite{Ref11}. Moreover, the stable manifold design is also preferable as mentioned by \cite{Ref12}. The non-linearity along with hybrid control and its application are broadly discussed in \cite{Ref13}, \cite{Ref14}, \cite{Ref15}, \cite{Ref16}, and \cite{Ref17}.

In this paper, the linearized and non-linear system is designed based on two different properties along with some further influences due to these variables. The normal state-space from three different characteristics of initial condition is designed in order to examine the properties of controllability and observability which are the key information of constructing the controller. The merged concept of state-feedback with canonical form from \cite{Ref18}, \cite{Ref19}, \cite{Ref20}, \cite{Ref21}, \cite{Ref22}, and \cite{Ref23} is then introduced to obtain the best value of $K$ leading to $K_d$ which is then simulated to perform the system. The influence of time sampling is used to examine the potency of degrading performance both in continuous- and discrete-system. The structure of this paper is started by introduction regarding the important of studying the inverted pendulum along with some development and constraints in first and the second chapter in turn. The third stage comprises the mathematical model dealing with the whole material while the following constitutes the illustrative example. The last is the conclusion being ended by some acknowledgement.

\section{Problem Formulation}
\label{}
This paper presents a mathematical modelling on inverted pendulum as shown in Fig.1 with the following qualitative objectives:
\begin{enumerate}\setlength\itemsep{0em}
    \item The problem is initiated from showing the right state-space in terms of nominally circular solution by considering small deviations of the state variables;
    \item Those small deviations are measured in terms of the small $|\phi(t)| < \frac{\pi}{4}, \forall t \geq 0$, the variables of $\dot{\phi}$ and $\dot	s$ should be maintained from excessively large $\forall t \geq 0$; and
    \item The states are derived with different initial condition with respect to the design of gain $K$
\end{enumerate}

\section{Mathematical Models}
\label{}
From the proposed aims, the foundation lies in the mathematical design of inverted pendulum along with the core feedback system. The followings are to design those two key parameters namely state-space and its properties along with the state-feedback.

\subsection{Design of Inverted Pendulum}
Analytical scheme in Fig.\ref{Fig 1} is deeply explained in order to obtain the physical system being divided into two directions, horizontal and vertical. Keep in mind that the centre of gravity is coordinated as $(\mathcal{S}_x, \mathcal{S}_y)$ in the pendulum rod. As for the forces, summing the the free-body diagram of the cart in the horizontal force axis is defined as follow:
\begin{align}
	M \frac{\partial^2}{\partial t^2} s(t) + m \frac{\partial^2}{\partial t^2} \mathcal{S}_x(t) + F \dot{s} = \mu(t) &&
    \label{Eq 1}
\end{align}
where the elaboration of the centre with respect to the distant $s$ and angle $\phi$ is described as,
\begin{align}
	\mathcal{S}_x = s + L \sin \phi \qquad \mathcal{S}_y = L \cos \phi &&
    \label{Eq 2}
\end{align}
from Eq.(\ref{Eq 1}) and (\ref{Eq 2}), it is rewarding to understand the basis concept of differential which then affects the formula. As regards the $(\sin)$ scenario, the first and the second derivative is shown using the formula $u'v + uv'$ as in the following,
\begin{align}
	\frac{\partial \sin \phi}{\partial t} &= \dot{\phi} \cos \phi  \nonumber\\ \frac{\partial^2 \sin \phi}{\partial t^2} &= \ddot{\phi} \cos \phi - \dot{\phi}^2 \sin \phi &&
    \label{Eq 3}
\end{align}
with respect to the $(\cos)$, the derivative concept is exactly the same as that of in $(\sin)$ as portrayed below,
\begin{align}
	\frac{\partial \cos \phi}{\partial t} &= - \dot{\phi} \sin \phi \nonumber \\ \frac{\partial^2 \cos \phi}{\partial t^2} &= - \ddot{\phi} \sin \phi - \dot{\phi}^2 \cos \phi &&
    \label{Eq 4}
\end{align}
as for the defined variables on Fig.\ref{Fig 1}, $M$ and $m$ is the mass of the carriage (cart) and the pendulum respectively while $L$ is the length of the pendulum. $s(t)$ is the displacement of the pivot in the $x-$axis and $\phi(t)$ is the angular rotation of the pendulum whereas $\mu(t)$ is the external force exerted on the carriage. In the bottom of the cart, there is a friction with certain coefficient and the system is influenced by the gravitational acceleration. According to Eq.(\ref{Eq 3}) and (\ref{Eq 4}), the formulation of the whole system in Eq.(\ref{Eq 1}) becomes:
\begin{figure}[t!]
	\centering
	\captionsetup{justification=centering}
	\begin{subfigure}[t]{\linewidth}
		\centering
		\raisebox{-0.5\height}{\includegraphics[scale=0.45]{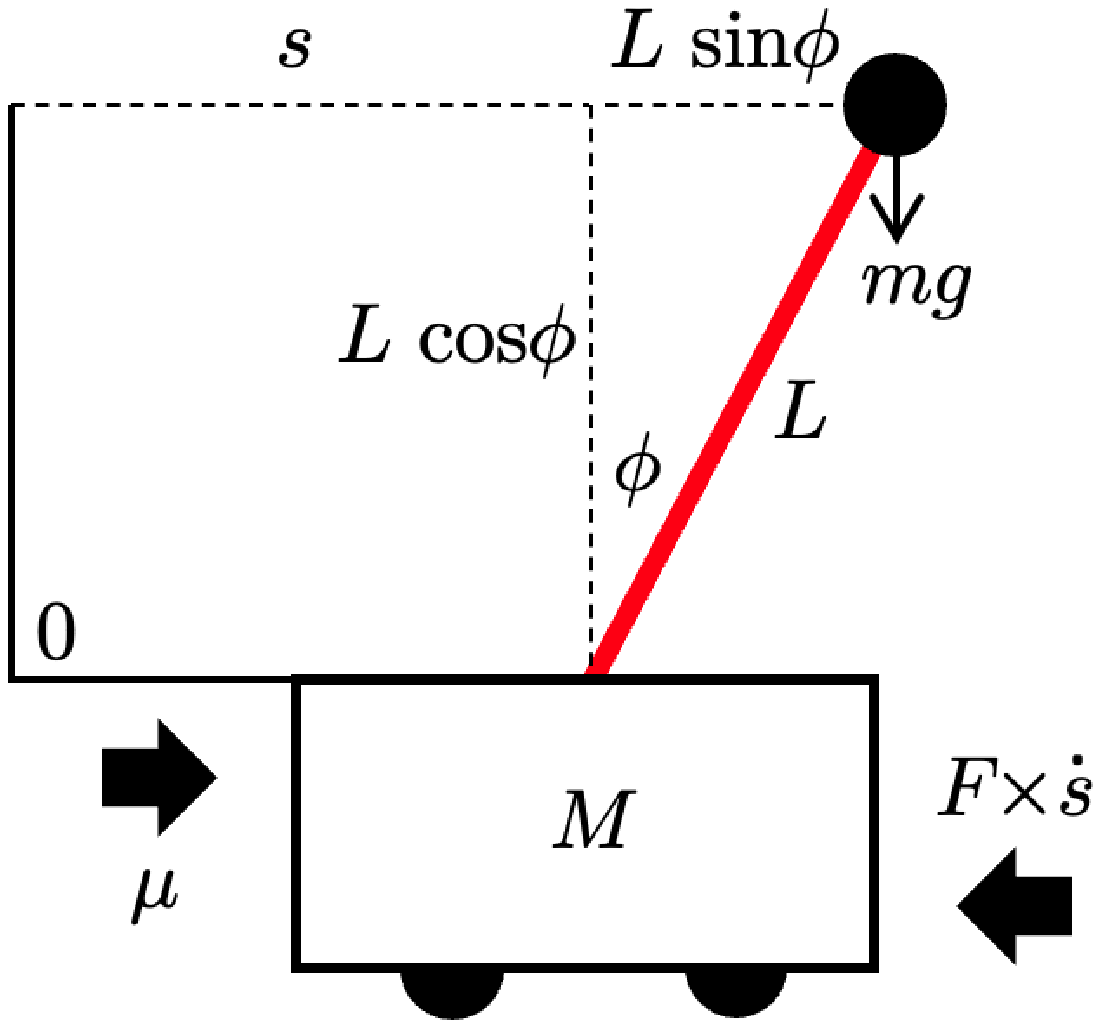}}
		\caption{}
		\label{Fig 1}
	\end{subfigure}
	\\
    \begin{subfigure}[t]{\linewidth}
		\centering
		\raisebox{-0.5\height}{\includegraphics[scale=0.45]{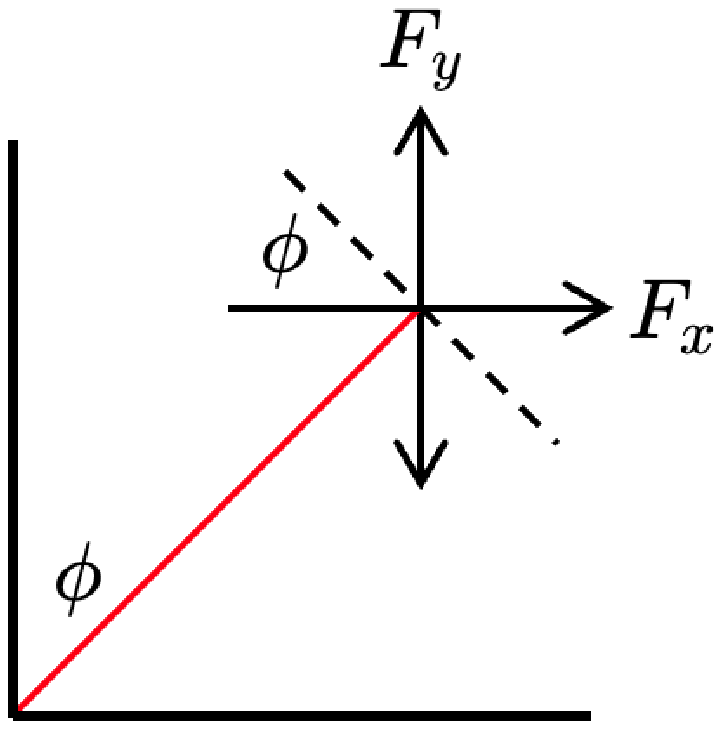}}
		\caption{}
		\label{Fig 2}
	\end{subfigure}
	\caption{(a) Inverted Pendulum on a cart, (b) Force components in torque balance}
\end{figure}
\begin{align}
	(M + m)\ddot{s} + mL \ddot{\phi} \cos \phi - mL \dot{\phi}^2 \sin \phi + F \dot{s} = \mu(t) &&
    \label{Eq 5}
\end{align}
suppose that if the point mass is relatively small, it can be ignored meaning that the cart contains only the rod without any concern to the rope. Moreover, assume that the there is no slip in the hinge between pendulum and the carriage so that the equation will be shown as follows:
\begin{align}
    M \ddot{s} + F \dot{s} - \mu(t) = 0 &&
    \label{Eq 6}
\end{align}

Moving to another direction so as to gain the second equation of this system, summing the forces being perpendicular to the pendulum relates to the torque balance on the system as depicted in Fig. \ref{Fig 2}. In this condition, the torque on the mass due to the acceleration and the gravity force are balancing so that the elaborated equation can be described in the following, such that:
\begin{align}
	F_x L \cos \phi - F_y L \sin \phi = mgL \sin \phi &&
    \label{Eq 7}
\end{align}
where $F_x$ and $F_y$ equal to
\begin{align}
    F_x \coloneqq &\; ma_x = m\left(\ddot{s} + L\ddot{\phi} \cos \phi - L\dot{\phi}^2 \sin \phi\right) \label{Eq 8}\\
    F_y \coloneqq &\; ma_y = m\left(-L\ddot{\phi} \sin \phi - L\dot{\phi}^2 \cos \phi\right) && \label{Eq 9}
\end{align}
with,
\begin{align*}
    a_x = \frac{\partial^2}{\partial t^2}\mathcal{S}_x \qquad \textrm{and} \qquad a_y = \frac{\partial^2}{\partial t^2}\mathcal{S}_y &&
\end{align*}
the combination of the formulas in the Eq.(\ref{Eq 7}), (\ref{Eq 8}) and (\ref{Eq 9}) generates another further equation as shown below, therefore:
\begin{align}
	\ddot{\phi} + \frac{1}{L} \ddot{s} \cos \phi - \frac{g}{L} \sin \phi = 0 &&
    \label{Eq 10}
\end{align}

\subsection{State-Space and Its Properties}
The findings in  Eq.(\ref{Eq 6}) results in acceleration $\ddot{s}$ which is applied in producing the angular acceleration $\ddot{\phi}$ variable from Eq. (\ref{Eq 10}). The general non-linear state-space representation is summed up in the following: 
\begin{align}
	\frac{d}{dt}x &= \overbrace{\mathcal{A}x + \mathcal{B}u + \xi_p}^{f(x,u,v)} \rightarrow (\textbf{E}, \mathcal{Q}) \sim x, \xi\label{Eq 11}\\
    y &= \underbrace{\mathcal{C}x + \mathcal{D}u + \eta_m}_{g(x,u,v)} \rightarrow (\textbf{E}, \mathcal{Q}) \sim y, \eta\label{Eq 12} &&
\end{align}
where the details of the states in the non-linear state-space are presented in the following, such that
\begin{align*}
    \dot{x}_1 =\; & x_2, \quad \dot{x}_2 = -\frac{F}{M}x_2 + \frac{1}{M}u, \quad \dot{x}_3 = x_4\\
    \dot{x}_4 =\; & \frac{g}{L}\sin(x_3) + \frac{F}{ML}\cos(x_3)x_2 - \frac{1}{ML}\cos(x_3)u\\
    y_1 =\; & x_1, \quad y_2 = x_3 &&
\end{align*}
with $\xi$ and $\eta$ denote the noise in the proses and measurement with certain Gaussian mean $\textbf{E}$ and covariance $\mathcal{Q}$ in turn. Standard format of the non-linear model of the inverted pendulum is illustrated by these Jacobian matrices, therefore:
\begin{align*}
    \dot{x} &\coloneqq \begin{bmatrix}
		\dot{x}_1\\[3pt]
		\dot{x}_2\\[3pt]
		\dot{x}_3\\[3pt]
		\dot{x}_4
    \end{bmatrix} = \begin{bmatrix}
    	\dot{s}\\[2pt]
		\ddot{s}\\[2pt]
		\dot{\phi}\\[2pt]
		\ddot{\phi}
    \end{bmatrix}, \qquad x \coloneqq \begin{bmatrix}
		x_1\\[3pt]
		x_2\\[3pt]
		x_3\\[3pt]
		x_4
    \end{bmatrix} = \begin{bmatrix}
    	s\\[2pt]
		\dot{s}\\[2pt]
		\phi\\[2pt]
		\dot{\phi}
    \end{bmatrix}, \\ u(t) &\coloneqq \mu(t), \qquad y(t) \coloneqq \begin{bmatrix}
    y_1\\[2pt]
    y_2
    \end{bmatrix} = \begin{bmatrix}
    s\\[2pt]
    \phi
    \end{bmatrix}  &&   
\end{align*}
suppose that there is no force on the system meaning that $\mu(t)$ equals to zero, the equilibrium points are represented by two conditions which are $\phi = 0$ (down position, stable) and  $\phi = \pi$  (up position, unstable). These values are computed based on a Taylor Series expansion, such that:
\begin{align}
	\mathcal{G}(\phi) = \mathcal{G}(\phi_n) + \phi \left.\frac{\partial \mathcal{G}}{\partial \phi}\right|_{\phi_n} \quad \longrightarrow \quad n = [0, \pi]
    \label{Eq 13} &&
\end{align}
with respect to \textbf{up}-position of equilibrium point $\phi = \pi$, the expansion of Taylor series in Eq.(\ref{Eq 13}) so as to receive the variables of $\cos \phi$, $\sin \phi$, and $\dot{\phi}^2$ for the sake of the states in Eq.(\ref{Eq 11}) and (\ref{Eq 12}), therefore:
\begin{align*}
    \mathcal{G}_{\phi = \pi} \simeq \begin{cases}
    \textrm{\small (1) }\cos \phi = \cos \pi + \left[(\pi - \phi)(-\sin \pi)\right] \\[10pt]
    \textrm{\small (2) }\sin \phi = \sin \phi + \left[(\pi - \phi)(\cos \pi)\right] \\[10pt]
    \textrm{\small (3) }\sin \phi = \pi - (\phi' + \pi) = - \phi'\\[10pt]
    \textrm{\small (4) }\dot{\phi}^2 \approx 0
    \end{cases} &&
\end{align*}
note: (1) and (2) result in $-1$ and $\pi - \phi$ respectively. From (2), analysis is simplified by defining a new coordinate, such that $\phi' = \phi - \pi$ and this is nothing more than a measured clockwise from the \textbf{up}-position, so that it is now (3). From the formula above, $\phi', \dot{\phi}', \ddot{\phi}'$ will be written as $\phi, \dot{\phi}, \ddot{\phi}$, so that this does not have any effects on the state equations. This linearization in terms of $\pi$ will affect the state of angular acceleration $\ddot{\phi}$ only when another down-equilibrium with $\phi = 0$ as follows, 
\begin{align*}
	\mathcal{G}_{\phi = 0} \simeq \begin{cases}
	\textrm{\small (1) }\cos \phi = \cos 0 - \phi \sin 0 \approx 1\\[10pt]
    \textrm{\small (2) }\sin \phi = \sin 0 + \phi \cos 0 \approx \phi\\[10pt]
    \textrm{\small (3) }\dot{\phi}^2 \approx 0
    \end{cases} &&
\end{align*}
those two set equilibrium $(0, \pi)$ lead to divergent state of matrix $A = J_A(x_0, u_0)$ although the $\phi = 0$ is applied instead. The whole state with influence of this circumstance is shown as follows:
\begin{align}
    \dot{x}_1 =\; & x_2, \quad \dot{x}_2 = -\frac{F}{M}x_2 + \frac{u}{M}\nonumber\\
    \dot{x}_3 =\; & x_4, \quad \dot{x}_4 = \frac{g}{L} + \frac{F}{ML}x_2 - \frac{u}{ML}\label{Eq 14}\\
    y_1 =\; & x_1, \quad y_2 = x_3\label{Eq 15} &&
\end{align}
in a trivial way, two matrices are classified as controllable if and only if the rank $(\rho)$ of the controllability matrix $\mathcal{C}_\lambda$ is exactly the same as that of in the $A$ matrix. Controllability matrix is defined as the combinations of both $A$ and $B$ matrix with $A$ as $n$ by $n$ matrix, such that:
\begin{align}
	\mathcal{C}_\lambda =& \begin{bmatrix}
	B & AB & \cdots & A^{n-1}B
	\end{bmatrix}, \quad \rho_\mathcal{C} = n \nonumber\\
	=& \begin{bmatrix}
	0 & \frac{1}{M} & \frac{-F}{M^2} & \frac{F^2}{M^3} \\[5pt]
	\frac{1}{M} & \frac{-F}{M^2} & \frac{F^2}{M^3} & \frac{-F^3}{M^4} \\[5pt]
	0 & \frac{-1}{ML} & \frac{F}{M^2L} & \frac{-F^2}{M^3L}-\frac{g}{ML^2} \\[5pt]
	\frac{-1}{ML} & \frac{F}{M^2L} & \frac{-F^2}{M^3L}-\frac{g}{ML^2} & \frac{F^3}{M^4L}+\frac{Fg}{M^2L^2} \end{bmatrix}
    \label{Eq 16} &&
\end{align}
observability $\mathcal{O}_\lambda$ has the same pattern as controllability in terms of the numbers of rank which has to be same as $A$ matrix. However, the arrangement of the observability matrix is different from the previous one. Instead of in the one row, observability matrix is arranged in the one column, such that: (suppose $A$ is $n$ by $n$ matrix)
\begin{align}
	\mathcal{O}_\lambda = \begin{bmatrix}
	C \\
    CA \\
    \vdots \\
    CA^{n-1}
	\end{bmatrix}, \quad \rho_\mathcal{O} = n
    \label{Eq 17} &&
\end{align}

\subsection{State-Feedback with Canonical Form}
Recalling the dynamics of the closed-loop system declared in Eqs. $(\ref{Eq 14})$ and $(\ref{Eq 15})$, keep in mind that the system poles are presented as $|sI - A| = 0$ being called as the characteristic roots. Since it is required to approach the zero-stability coming from divergent arbitrary initialization states, the stability properties must be achieved in the internal system, such that:
\begin{align*}
	u(t) = -Kx(t) &&
\end{align*}
and the desired poles should be set. Furthermore, having designed the state vector which is looped to the input again, it is important to track the reference signal, such that:
\begin{align}
	u(t) = -Kx(t) + \Psi \delta(t)
    \label{Eq 18} &&
\end{align}
from Eq. (\ref{Eq 18}), it is substituted to Eq. (\ref{Eq 11}) generating the new state which will be then compared as the normal system,
\begin{align}
	\dot{x} = (A - BK)x + B\Psi \delta
    \label{Eq 19}&&
\end{align}
the determinant of this equation should be matched as the normal system in terms of the coefficient from those two characteristic polynomials, such that $|sI - A| = |sI - (A - BK)| = 0$. Due to the high dimension of computing the determinant, it is absolutely not a trivial calculation so that the canonical form is applied to fit the value of $K$. Recalling the controllability matrix $\mathcal{C}$ and denoting the transformation matrix $\Phi$, it is defined as $\Phi = \mathcal{C} \, \Gamma$ with,
\begin{align}
	\Gamma = \begin{bmatrix}
	\gamma_{n-1} & \gamma_{n-2} & \cdots & \gamma_1 & 1\\
    \gamma_{n-2} & \gamma_{n-3} & \cdots & 1 & 0\\
    \vdots & \vdots & \ddots & \vdots & \vdots\\
    \gamma_1 & 1 & \cdots & 0 & 0\\
    1 & 0 & \cdots & 0 & 0
	\end{bmatrix}\label{Eq 20} &&
\end{align}
those $\gamma_i$ are derived from the extract from the determinant of normal system, such that:
\begin{align}
	|sI - A| = s^n + \gamma_1 s^{n - 1} + \cdots + \gamma_{n-1}s + \gamma_n	
    \label{Eq 21} &&
\end{align}
the canonical form of $\Phi$ is then used to generate the updated design of $\hat{A} = \Phi^{-1} A \Phi$ and $\hat{B} = \Phi^{-1} B$, such that:
\begin{figure*}[h!]
\begin{align}
	\hat{A} &= \begin{bmatrix}
	0 & 1 & 0 & \cdots & 0\\
    0 & 0 & 1 & \cdots & 0\\
    \vdots & \vdots & \vdots & \ddots & \vdots\\
    0 & 0 & 0 & \cdots & 1\\
    -\gamma_n & -\gamma_{n-1} & -\gamma_{n-2} & \cdots & -\gamma_1
	\end{bmatrix}, \quad \hat{B} = \begin{bmatrix}
	0\\
    0\\
    \vdots\\
    0\\
    1
	\end{bmatrix}\nonumber\\
	\hat{A} - \hat{B}\hat{K} &= \begin{bmatrix}
	0 & 1 & 0 & \cdots & 0\\
    0 & 0 & 1 & \cdots & 0\\
    \vdots & \vdots & \vdots & \ddots & \vdots\\
    0 & 0 & 0 & \cdots & 1\\
    -\left(\bar{\gamma}_n - \hat{k}_1\right) & -\left(\bar{\gamma}_{n-1} - \hat{k}_2\right) & -\left(\bar{\gamma}_{n-2} - \hat{k}_3\right) & \cdots & -\left(\bar{\gamma}_1 - \hat{k}_n\right)
	\end{bmatrix}
    \label{Eq 22}\\
    \left|sI - \left(\hat{A} - \hat{B} \hat{K}\right)\right| &= s^n + \left(\bar{\gamma}_1 + \hat{k}_n\right)s^{n - 1} + \cdots + \left(\bar{\gamma}_{n-1} + \hat{k}_2\right)s + \left(\bar{\gamma}_n + \hat{k}_1\right)
    \label{Eq 23}&&
\end{align}
\end{figure*}
The following scheme is to define $\hat{K}$ as $[\hat{k}_1, \hat{k}_2, \cdots, \hat{k}_n]$. Keep in mind that the properties of eigenvalues in $\hat{K}$ does not change under transformation of $\Phi$. The Eq.(\ref{Eq 21}) and (\ref{Eq 23}) need to compare in order to get the collection of $\hat{k}_i$, such that:
\begin{align*}
	\hat{k}_n = \gamma_1 - \bar{\gamma}_1, \; \cdots, \; \hat{k}_{n-1} = \gamma_2 - \bar{\gamma}_2, \; \hat{k}_1 = \gamma_n - \bar{\gamma}_n &&
\end{align*}
where the actual $K$ is designed as $K = \hat{K} \Phi^{-1}$. The equations enlightening the relations among the state-space matrices of continuous- and discrete-time by step response invariance are:
\begin{align}
    A_d = e^{AT}, \quad B_d = \int_0^T e^{A(T-\tau)} B\, d\tau, \quad C_d = C &&
\end{align}

\section{Simulation Results}
\label{}
\begin{figure*}[t!]
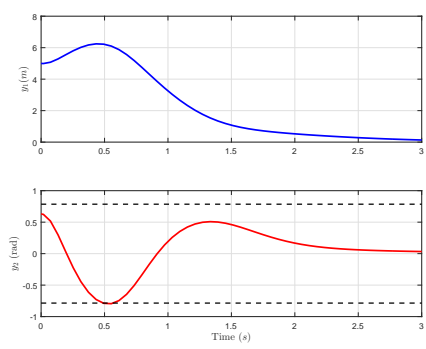
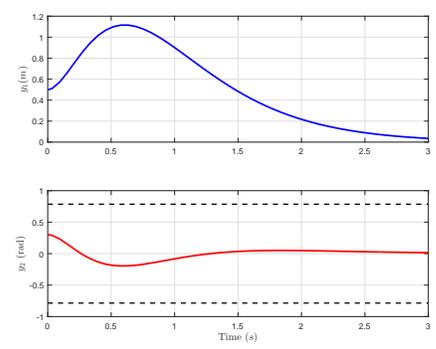

	\centering
	\captionsetup{justification=centering}
	\begin{subfigure}[t]{0.32\linewidth}
		\centering
		\raisebox{-0.1\height}{\includegraphics[width=.95\textwidth]{myplot1.eps}}
		\caption{Linearized, $x_u$, \textbf{C}}
		\label{Fig 3a}
	\end{subfigure}
	~
    \begin{subfigure}[t]{0.32\linewidth}
		\centering
		\raisebox{-0.1\height}{\includegraphics[width=.95\textwidth]{myplot2.eps}}
		\caption{Linearized, $x_c$, \textbf{C}}
		\label{Fig 3b}
	\end{subfigure}
	~
    \begin{subfigure}[t]{0.32\linewidth}
		\centering
		\raisebox{-0.1\height}{\includegraphics[width=.95\textwidth]{myplot3.eps}}
		\caption{Linearized, $x_s$, \textbf{C} $\And$ \textbf{D}}
		\label{Fig 3c}
	\end{subfigure}\\[7pt]
	\begin{subfigure}[t]{0.32\linewidth}
		\centering
		\raisebox{-0.5\height}{\includegraphics[width=.95\textwidth]{myplot4.eps}}
		\caption{Non-linear, $x_u$, \textbf{C}}
		\label{Fig 4a}
	\end{subfigure}
	~
    \begin{subfigure}[t]{0.32\linewidth}
		\centering
		\raisebox{-0.5\height}{\includegraphics[width=.95\textwidth]{myplot5.eps}}
		\caption{Non-linear, $x_c$, \textbf{C}}
		\label{Fig 4b}
	\end{subfigure}
	~
    \begin{subfigure}[t]{0.32\linewidth}
		\centering
		\raisebox{-0.5\height}{\includegraphics[width=.95\textwidth]{myplot6.eps}}
		\caption{Non-linear, $x_s$, \textbf{C} $\And$ \textbf{D}}
		\label{Fig 4c}
	\end{subfigure}\\[7pt]
	\begin{subfigure}[t]{0.32\linewidth}
		\centering
		\raisebox{-0.5\height}{\includegraphics[width=.95\textwidth]{myplot7.eps}}
		\caption{$x_s, \textbf{C}, T = 0.5$}
		\label{Fig 5a}
	\end{subfigure}
	~
    \begin{subfigure}[t]{0.32\linewidth}
		\centering
		\raisebox{-0.5\height}{\includegraphics[width=.95\textwidth]{myplot8.eps}}
		\caption{$x_s, \textbf{C}, T = 0.2$}
		\label{Fig 5b}
	\end{subfigure}
	~
    \begin{subfigure}[t]{0.32\linewidth}
		\centering
		\raisebox{-0.5\height}{\includegraphics[width=.95\textwidth]{myplot9.eps}}
		\caption{$x_s, \textbf{C}, T = 0.1$}
		\label{Fig 5c}
	\end{subfigure}\\
    \begin{subfigure}[t]{0.48\linewidth}
		\centering
		\raisebox{-0.5\height}{\includegraphics[width=.95\textwidth]{myplot13.eps}}
		\caption{$x_s$, \textbf{C}, Performance degradation due to $T$}
		\label{Fig 7a}
	\end{subfigure}
	~
    \begin{subfigure}[t]{0.48\linewidth}
		\centering
		\raisebox{-0.5\height}{\includegraphics[width=.95\textwidth]{myplot14.eps}}
		\caption{$x_s$, \textbf{D}, Performance degradation due to $T$}
		\label{Fig 7b}
	\end{subfigure}
	\caption{Performance variations of \textbf{C} (Continuous) and \textbf{D} (Discrete) from linearized and non-linear system with the influences of three different characteristics of initial conditions $x_u, x_c, x_s$ along with some gain of $K, K_d$ and time-sampling $T$}
\end{figure*}
To illustrate the scheme  along with the proposed state-feedback with canonical form, the followings variables are designed, such as $M = 1$kg, $L = 0.842$m, $F = 1$kg.s$^{-1}$, and $g = 9.8093$m.s$^{-2}$. Suppose that $x_1 = s$, $x_2 = \dot{s}$, $x_3 = \phi$, and $x_4 = \dot{\phi}$ so that the matrix of $A$ and $B$ will be,
\begin{align*}
	A = \begin{bmatrix}
	0 & 1 & 0 & 0\\
    0 & -1 & 0 & 0\\
    0 & 0 & 0 & 1\\
    0 & 1.1876 & 11.6500 & 0
	\end{bmatrix} \quad B = \begin{bmatrix}
	0\\
    1\\
    0\\
    -1.1876
	\end{bmatrix} &&
\end{align*}
in terms of the controllability, the pair $(A,C)$ is controllable because it has a full rank of $\rho_\mathcal{C} = n = 4$ and it is also observable with the same rank $\rho_\mathcal{O} = n$, therefore
\begin{align*}
    \mathcal{C} &= \begin{bmatrix}
	0 & 1 & -1 & 1\\
   1 & -1 & 1 & -1\\
    0 & -1.1876 & 1.1876 & -15.0238\\
    -1.1876 & 1.1876 & -15.0238 & 15.0238
	\end{bmatrix} &&
\end{align*}
\begin{align*}
	\mathcal{O} &= \begin{bmatrix}
    1 & 0 & 0 & 0\\
    0 & 0 & 1 & 0\\
    0 & 1 & 0 & 0\\
    0 & 0 & 0 & 1\\
    0 & -1 & 0 & 0\\
    0 & 1.1876 & 11.6500 & 0\\
    0 & 1 & 0 & 0\\
    0 & -1.1876 & 0 & 11.6500
   \end{bmatrix} &&
\end{align*}
having computed the analytical problems, initial some constraints are delivered, such that:
\begin{align*}
	|\phi| < \frac{\pi}{4}; \quad |\dot{\phi}(t)| \textmd{ and } |\dot{s}(t)| \neq \gg \textrm{ large} \; \forall \, t \geq 0 &&
\end{align*} and
\begin{align*}
    x_u &= \begin{bmatrix}7 & 0 & \frac{\pi}{2} & 0\end{bmatrix} \qquad x_c = \begin{bmatrix}5 & -1 & \frac{\pi}{5} & 0.2\end{bmatrix}, \\
	x_s &= \begin{bmatrix}0.5 & 0 & 0.3 & 0\end{bmatrix} &&
\end{align*}along with
\begin{align*}
    \mathcal{P} &= \begin{bmatrix}-2 & -3+0.5i & -3-0.5i & -4\end{bmatrix}, \\
	\mathcal{P}_d &= \begin{bmatrix}0.61 & 0.47+0.06i & 0.47-0.06i & 0.37\end{bmatrix} &&
\end{align*}
where the initial conditions of $x_u, x_c, x_s$ refer to unstable (large displacement and angular rotation), critical and stable (equilibrium) in turn. Since the open-loop linearized system is unstable because its poles are not strictly in the negative real part of complex plane, the poles of the state-feedback is proposed as $\mathcal{P}$. Theoretically, the large negative poles lead to the fast stability yet yielding a large gain of $K$, making the design seems unrealistic or poor to the true systems.

Furthermore, those initial conditions are also implemented in the linearized as shown in Fig.(\ref{Fig 3a}), (\ref{Fig 3b}), (\ref{Fig 3c}) and non-linear system showing that the large displacement and angular rotations made the control unstable. With these angles, the linearisation is no longer an acceptable approximation of the real system, thus the control designed for the linear system is not working properly as depicted in Fig.(\ref{Fig 4a}), (\ref{Fig 4b}), (\ref{Fig 4c}). The control law implementation with a discrete-time controller is applied only for the stable $x_s$ initial condition as portrayed in Fig.(\ref{Fig 5a}), (\ref{Fig 5b}), (\ref{Fig 5c}). The opted time sampling is also crucial and the stability is achieved with $T = 0.1$. At $T \geq 1s$. The main observation is that the behaviour of the nonlinear system with sampling is also dependant on the chosen initial conditions. An adjustment of the gain with respect to the sampling time would be essential in order to stabilize the system for higher sampling time values in discrete time  as depicted in Fig.(\ref{Fig 7a}). In order to perform a fair comparison between the continuous-time and discrete-time models, the chosen desired poles were transformed from the $s$-plane to the $z$-plane with $z = e^{sT}$. This transformation guarantees that the poles of the discretized linearised system are inside the unity disk, thus the stability of the system is preserved. This is because the poles on the s-plane with negative real part are mapped inside the unity disk on the $z$-plane. Moreover, due to the pole correspondence between the $s$- and $z$-plane, the performance of the discrete-time is expected to be similar to the continuous-time system
\section{Conclusion}
Increasing the sampling time (hence recalculating the discrete poles and matrices for each sampling time values) has influence on the performance of the system. The peak value of $y_1$ is increasing and after some value of $T$ an overshoot starts to be present as $T$ increases. With the redesigned controller $K_d$, stability can be achieved for larger sampling time values than in the case of $K$, but the value of $T$ has to be chosen such a way that sufficient amount of information of the system can be obtained. The smaller the sampling time, the closer performance can be achieved to the continuous time case. On the contrary increasing the sampling time leads to degrading dynamical performance, and at last, instability

\bibliographystyle{IEEEtran}
\bibliography{reference.bib}

\end{document}